\documentclass[a4paper]{jpconf}
\usepackage{graphicx}
\usepackage{amsfonts}
\usepackage{epsfig}
\usepackage{color}
\usepackage{multirow}

\usepackage{lipsum}					
\usepackage[centertags]{amsmath}			
\usepackage{stmaryrd}					
\usepackage{amssymb}					
\usepackage{amsthm}					
\usepackage{slashed}
\usepackage{layouts}					
\usepackage{graphicx}					
\usepackage{subfig}
\usepackage{longtable,rotating}			
\usepackage{colortbl}					
\usepackage{float}						
\usepackage{verbatim}					
\usepackage{upgreek }					
\usepackage{textgreek}
\usepackage[square,numbers,
		     sort&compress]{natbib}		
\usepackage{url}						
\usepackage[spanish,english,french]{babel}		
\usepackage{color}                    				
\usepackage[colorlinks=true,
		     allcolors=black]{hyperref}              
\usepackage{memhfixc}					
\usepackage{microtype}					
\usepackage{rotfloat}					
\usepackage{alltt}						
\usepackage[version=0.96]{pgf}			
\usepackage{tikz}						
\usetikzlibrary{arrows,shapes,snakes,
		       automata,backgrounds,
		       petri,topaths}				
\usepackage{feynmp-auto}
\usepackage{import}
\usepackage{calc,soul,fourier}

\newcommand{\be}{\begin{equation}}
\newcommand{\ee}{\end{equation}}
\newcommand{\ba}{\begin{eqnarray}}
\newcommand{\ea}{\end{eqnarray}}

\newcommand{\unit}{\mathbb{I}}
\newcommand{\nn}{\nonumber}

\newcommand{\pgftextcircled}[1]{                                                                    
    \setbox0=\hbox{#1}%
    \dimen0\wd0%
    \divide\dimen0 by 2%
    \begin{tikzpicture}[baseline=(a.base)]%
        \useasboundingbox (-\the\dimen0,0pt) rectangle (\the\dimen0,1pt);
        \node[circle,draw,outer sep=0pt,inner sep=0.1ex] (a) {#1};
    \end{tikzpicture}
}
\newcommand{\blackged}{\hfill$\blacksquare$}
\newcommand{\whiteged}{\hfill$\square$}

\newcounter{proofcount}

\renewcommand{\thefootnote}{\arabic{footnote}} 	
\begin{document}

\title{The Elastic $q\bar q$ Cross Section in the Nambu--Jona-Lasinio Model}

\author
{Rafael Chapelle$^1$, Joerg Aichelin$^{1,2}$ , Juan M. Torres-Rincon$^2$}
\address{$^1$ Subatech, UMR 6457, IN2P3/CNRS, Universit\'e de Nantes, \'Ecole de Mines de Nantes, 4 rue Alfred Kastler 44307,
Nantes, France}
\address{$^2$ Frankfurt Institute for Advances Studies. Johann Wolfgang Goethe University, Ruth-Moufang-Str. 1,
60438, Frankfurt am Main, Germany}

\begin{abstract}
We discuss the quark masses and the elastic  $q\bar q$ cross sections at finite chemical potential in the Nambu--Jona-Lasinio model. 
We comment the generic features of the cross sections as functions of the chemical potential, temperature and collision energy. Finally, we discuss their 
relevance in the construction of a relativistic transport model for heavy-ion collisions based on this effective Lagrangian.
\end{abstract}

\section{Introduction}
 The Nambu--Jona-Lasinio (NJL) model
has been extensively used in the context of strong interactions due to its ability to account for several key
phenomena of the Quantum Chromodynamics (QCD), like the spontaneous symmetry breaking (together with the generation of
Goldstone bosons) and its restoration at high temperatures and densities. This model works as an effective realization of 
QCD at low energies, and allows us performing studies in a much simpler way in the regime where QCD is too difficult
to solve, or computationally expensive like in lattice-QCD calculations. This fact we use here to derive elastic $q\bar q$ cross sections in
the non-perturbative region. These cross sections are needed if one wants to simulate the expansion of a plasma created in ultrarelativistic
heavy-ion collisions.

We consider the Lagrangian of the NJL model with (color neutral) pseudoscalar and scalar interactions (neglecting the vector and axial-vector vertices for simplicity)~\cite{Torres-Rincon},
\ba 
 {\cal L}_{NJL} &=& \sum_i \bar{\psi}_i (i \slashed{\partial}-m_{0i}+\mu_{i} \gamma_0) \psi_i \nn \\
&+& G \sum_{a} \sum_{ijkl} \left[ (\bar{\psi}_i \ i\gamma_5 \tau^{a}_{ij} \psi_j) \ 
(\bar{\psi}_k \ i \gamma_5 \tau^{a}_{kl} \psi_l)
+ (\bar{\psi}_i \tau^{a}_{ij} \psi_j) \ 
(\bar{\psi}_k  \tau^{a}_{kl} \psi_l) \right] \nn \\
& -&    K \det_{ij} \left[ \bar{\psi}_i \ ( \unit - \gamma_5 ) \psi_j \right] - K \det_{ij} \left[ \bar{\psi}_i \ ( \unit + \gamma_5 ) \psi_j \right]   
\label{eq:lagPNJL}
\ea
where the flavor indices $i,j,k,l=1,2,3$ and $\tau^{a}$ ($a=1,...,8$) being the $N_f=3$ flavor generators with
normalization 
\be \textrm{tr}_f \  (\tau^{a} \tau^{b}) = 2\delta^{ab}  \ , \ee
with $\textrm{tr}_f$ denoting the trace in flavor space.  In the Lagrangian~(\ref{eq:lagPNJL}) the bare quark masses are represented by $m_{0i}$ and their chemical potential by
$\mu_{i}$. The coupling constant for the scalar and pseudoscalar interaction $G$ is taken as
a free parameter (fixed e.g. by the pion mass in vacuum). The third term of Eq.~(\ref{eq:lagPNJL}) is the so-called 't Hooft Lagrangian, which mimics the effect of the axial $U(1)$ anomaly, accounting for
the physical splitting between the $\eta$ and the $\eta'$ meson masses. $K$ is an unknown coupling constant (fixed by the value
of $m_{\eta'}-m_{\eta}$) and $\unit$ is the identity matrix in Dirac space. As the NJL model is non-renormalizable, it also requires an ultraviolet regulator, which we introduce in the form of a cutoff $\Lambda$.

This Lagrangian has been widely used to study strongly interacting systems in the vacuum and at finite temperature. It contains 5 parameters. For all details we refer to the reviews~\cite{Vogl,Klevansky,Hatsuda,Alkofer,Buballa}.

\section{Masses of $u$, $d$ and $s$ quarks as functions of temperature and chemical potential}
In the SU(3) version of the NJL Lagrangian the mass of a quark of flavor $i$ is given by 
\begin{equation}
m_i = m_{0i} - 4G \langle \overline{q}_i  q_i \rangle + 2 K \langle \overline{q}_j q_j \rangle \langle \overline{q}_k  q_k \rangle \ ,
\label{Eq:gap}
\end{equation}
where $m_{0i}$ is the bare mass,  $i \neq j \neq k$ are the $u$, $d$, $s$ quarks, $ \langle \overline{q}_i  q_i \rangle$ the scalar condensate which, with the thermal distribution function
\begin{equation}
f^\pm_i(E_i,T,\mu_i) = \frac{1}{1+\exp \left( \frac{E_i \mp \mu_i}{T} \right)} \ ,
\end{equation}
is given by 
\begin{equation}
\langle \bar{q}_i q_i \rangle= - 2 N_c  \int_0^\Lambda \frac{m_i}{\sqrt{p^2+m_i^2}}(1-f_i^{+}-f_i^{-}) \frac{\mathrm{d}^3p}{(2 \pi)^3} \ .
\end{equation}
\begin{figure}[htp]%
\centering
\includegraphics[scale=0.5]{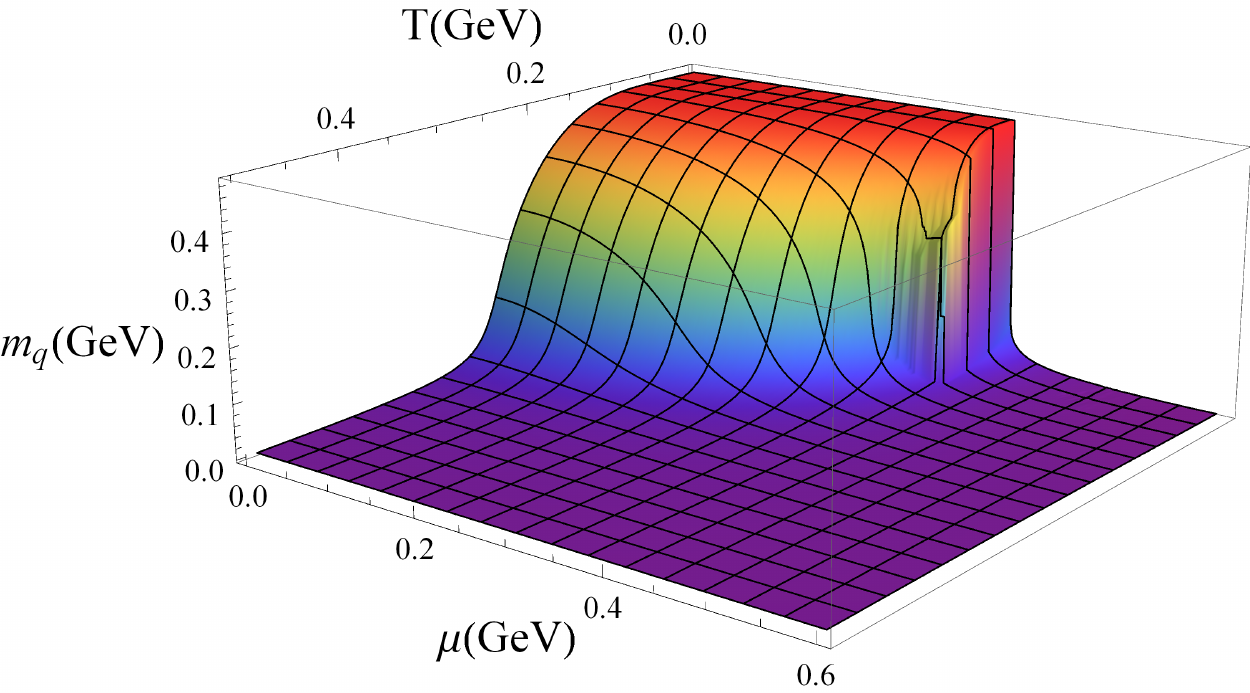}
\includegraphics[scale=0.5]{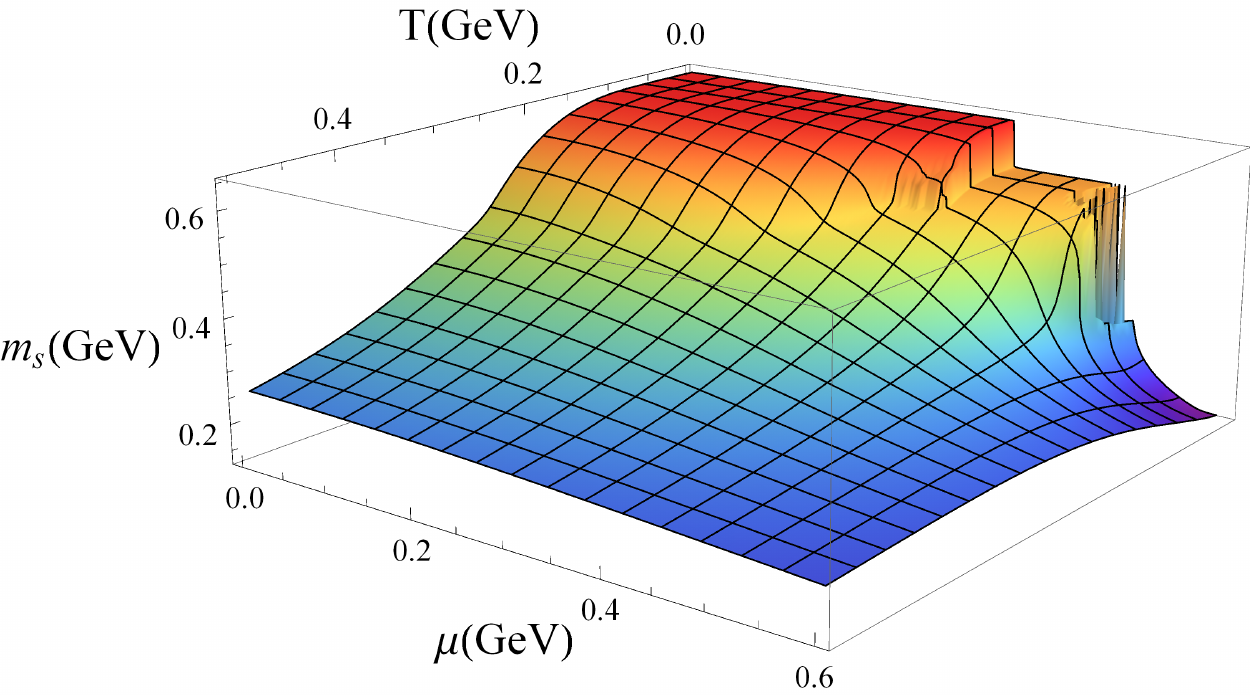}%
\caption{\label{fig:masses}The mass of $u$, $d$ (left) and the $s$ (right) quark as functions of the temperature and the chemical potential.}
\end{figure}
 In the present calculation we employ the parameters of Table~\ref{tab:para}, determined by vacuum meson masses and decay constants.
\begin{center}
\begin{table}[h]
\caption{\label{tab:para}Parameters used in the calculation and the critical chemical potential $\mu_{crit}$, where the chiral restoration occurs as a first order phase transition.}
\centering
\begin{tabular}{|c | c | c | c| c | c |}
\hline
$\Lambda$ & $G$ & $K$ & $m_{0u}$ & $m_{0s}$ & $\mu_{crit}$ \\ \hline
569 MeV & $2.3/\Lambda^2$  & $11/\Lambda^5$ & 5.5 MeV & 134 MeV & 338 MeV\\ \hline
\end{tabular}
\end{table}
\end{center}
For $\mu = 0$ ($\mu$ denotes the light chemical potential, $\mu_u=\mu_d$) we observe for all quarks a smooth transition of the mass as a function of the temperature.  This behavior continues for $u$ and $d$ quarks
along the transition line between the quark plasma and hadrons until $\mu_{crit}$ is reached. Then the crossover becomes a first order phase transition and we see a sudden change of the masses. For the $s$
quark the situation is more complicated. We see a first sudden but continuous change at the phase transition for the $u$ and $d$ quarks because both are related (Eq.~\ref{Eq:gap}) and then a second first order transition when the $s$
quark mass become discontinuous. 

\section{The $q\bar q$ cross sections}
The  $q\bar q$ cross section can be obtained in the standard way from the Lagrangian~\cite{Rehberg}. The Feynman diagrams for the two channels
\let\textcircled=\pgftextcircled
\begin{figure}[htp]%
\centering
\begin{minipage}{0.49\textwidth}
\centering
\begin{fmffile}{simple1}
\begin{fmfgraph*}(140,90)
\fmfleft{i1,i2}
\fmfright{o1,o2}
\fmflabel{$p_2,m_2$}{i1}
\fmflabel{$p_1,m_1$}{i2}
\fmflabel{$p_4,m_4$}{o1}
\fmflabel{$p_3,m_3$}{o2}
\fmf{fermion,label}{i1,v1}
\fmf{fermion,label}{v1,o1}
\fmf{dbl_plain,label=$p=(p_1 - p_3)$,l.s=left}{v1,v2}
\fmf{fermion}{o2,v2}
\fmf{fermion}{v2,i2}
\end{fmfgraph*}
\end{fmffile}
\end{minipage}
\hfill
\begin{minipage}{0.49\textwidth}
\centering
\begin{fmffile}{simple3}
\begin{fmfgraph*}(140,90)
\fmfleft{i1,i2}
\fmfright{o1,o2}
\fmflabel{$p_2,m_2$}{i1}
\fmflabel{$p_1,m_1$}{i2}
\fmflabel{$p_4,m_4$}{o1}
\fmflabel{$p_3,m_3$}{o2}
\fmf{fermion}{i1,v1,i2}
\fmf{dbl_plain,label=$p=(p_1+p_2)$}{v1,v2}
\fmf{fermion}{o1,v2,o2}
\end{fmfgraph*}
\end{fmffile}
\end{minipage}
\\[1.0cm]
\caption{$t-$ and $s-$channel Feynman diagrams for elastic $q\bar q$ collisions.}
\label{fig:Diagfqq}
\end{figure}
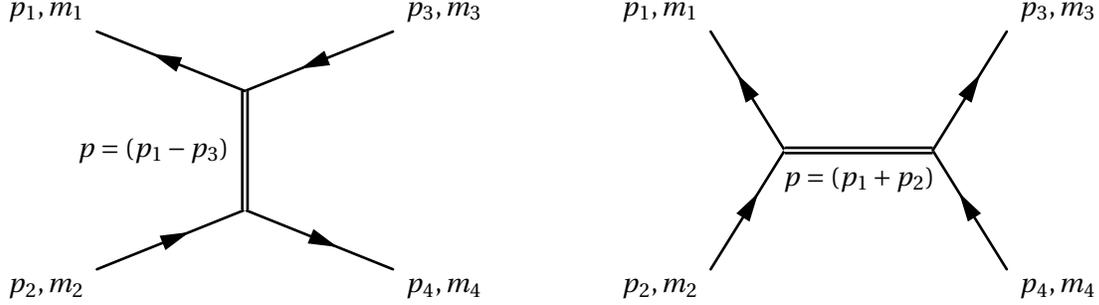
lead to the following matrix elements, 
\begin{equation}
\begin{split}
-i \mathcal{M}_t &= \delta_{c1,c3} \delta_{c2,c4} \bar{u}(p_3) T v(p_1) [i \mathcal{D}_t^S(p_1 - p_3)]\bar v(p_2) T  u(p_4)\\
& + \delta_{c1,c3} \delta_{c2,c4} \bar{u}(p_3)(i \gamma_5 T) v(p_1) [i \mathcal{D}_t^P(p_1 - p_3)]\bar{v}(p_2) (i \gamma_5 T) u(p_4) \ ,
\end{split}
\label{Eq:4.1}
\end{equation}
\begin{equation}
\begin{split}
-i \mathcal{M}_s &= \delta_{c1,c2} \delta_{c3,c4} \bar{v}(p_2) T u(p_1) [i \mathcal{D}_s^S(p_1 + p_2)]\bar{u}(p_3) T v(p_4)\\
& + \delta_{c1,c2} \delta_{c3,c4} \bar{v}(p_2)(i \gamma_5 T) u(p_1) [i \mathcal{D}_s^P(p_1 - p_4)]\bar{u}(p_3) (i \gamma_5 T) v(p_4) \ .
\end{split}
\label{Eq:4.3}
\end{equation}
\begin{figure}[htp]%
\centering
\begin{minipage}{0.49\textwidth}
\centering
\includegraphics[scale=0.5]{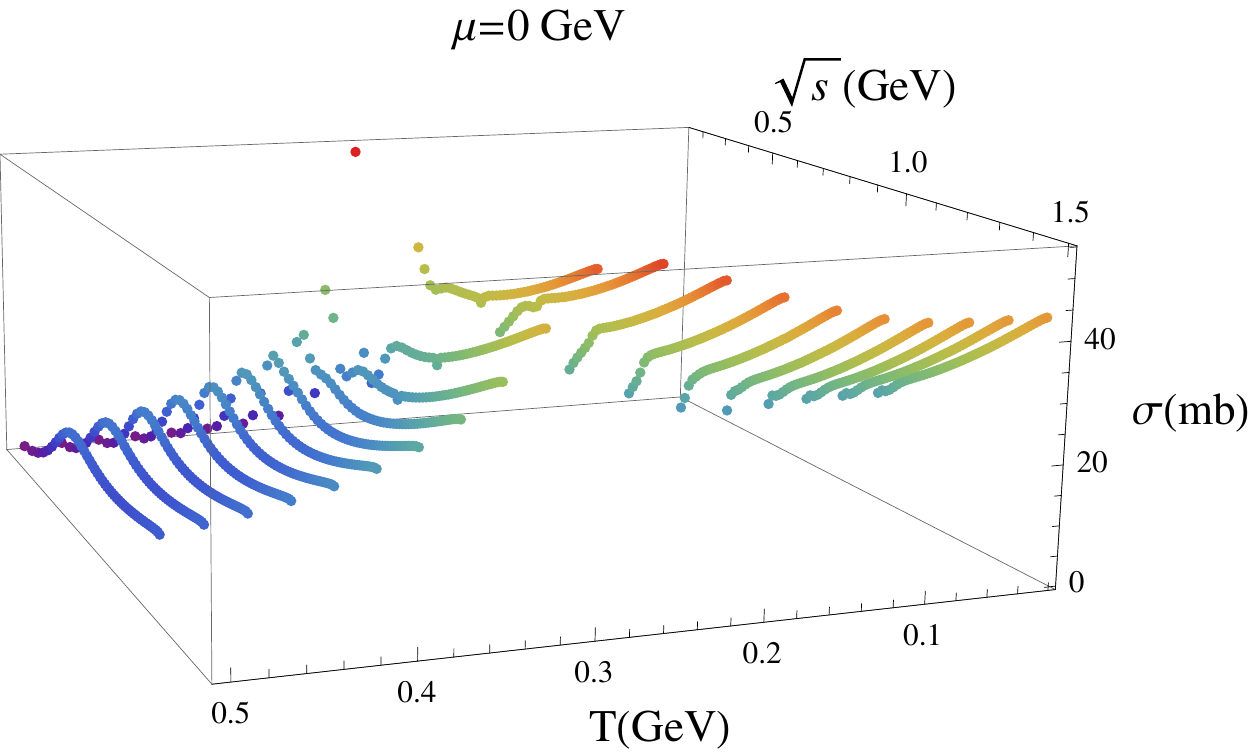}
\end{minipage}
\hfill
\begin{minipage}{0.49\textwidth}
\includegraphics[scale=0.5]{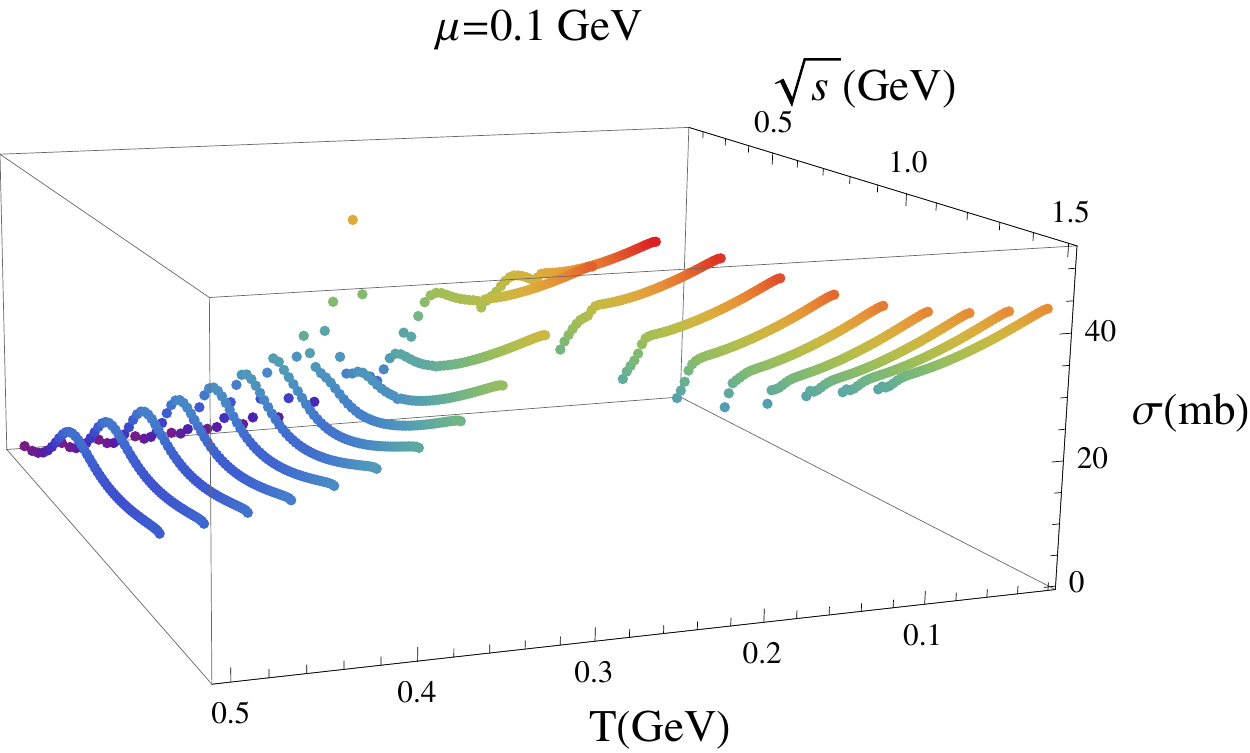}
\end{minipage}
\hfill
\begin{minipage}{0.49\textwidth}
\includegraphics[scale=0.5]{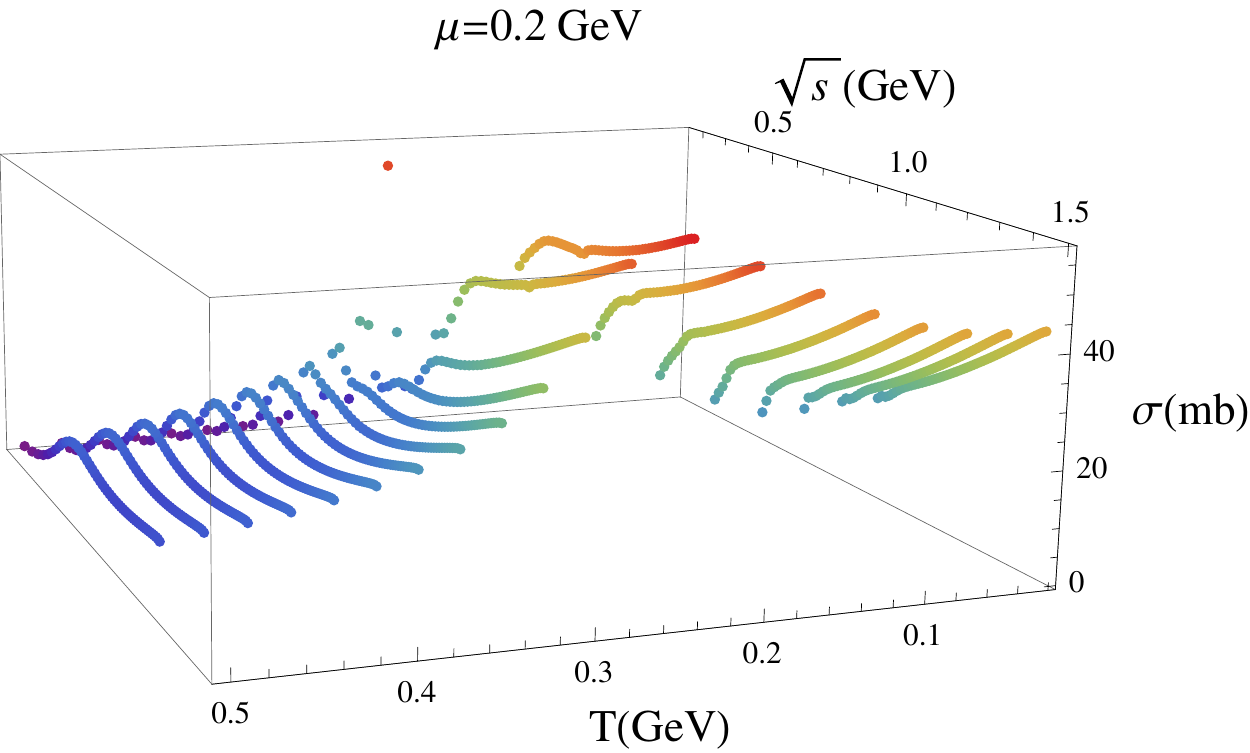}
\end{minipage}
\hfill
\begin{minipage}{0.49\textwidth}
\includegraphics[scale=0.5]{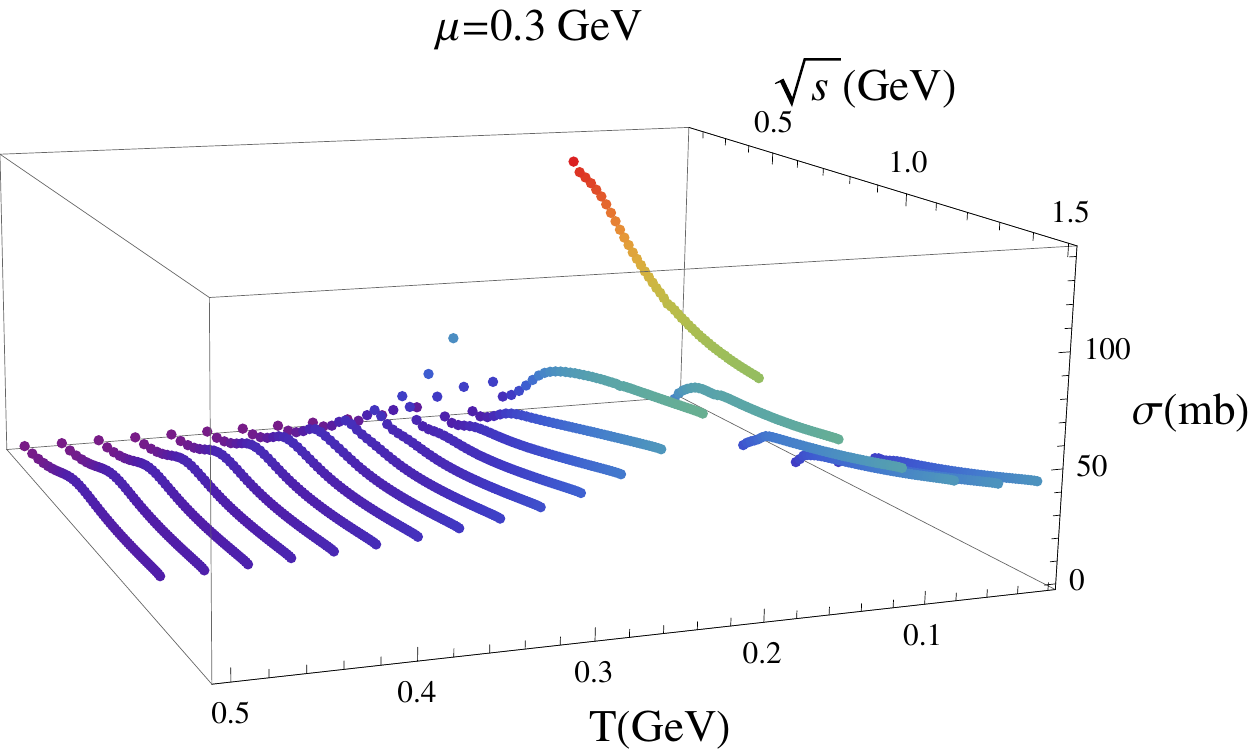}
\end{minipage}
\hfill
\begin{minipage}{0.49\textwidth}
\includegraphics[scale=0.5]{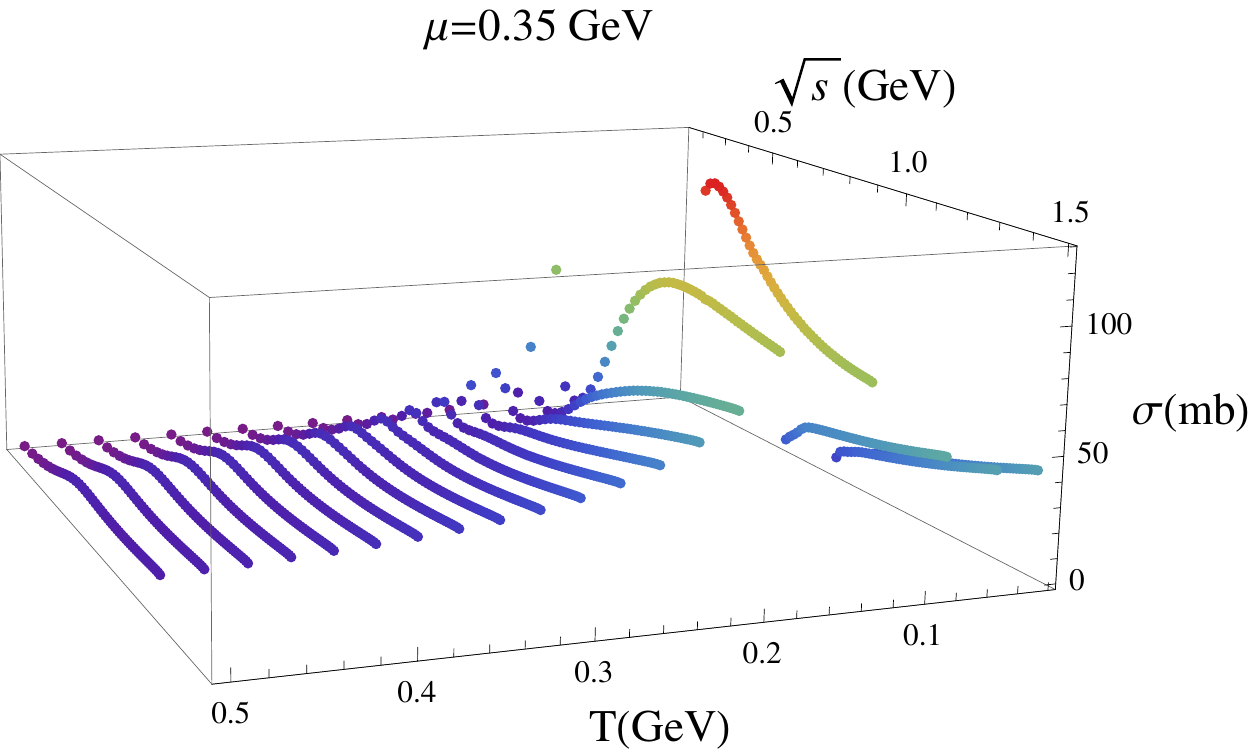}
\end{minipage}
\begin{minipage}{0.49\textwidth}
\includegraphics[scale=0.5]{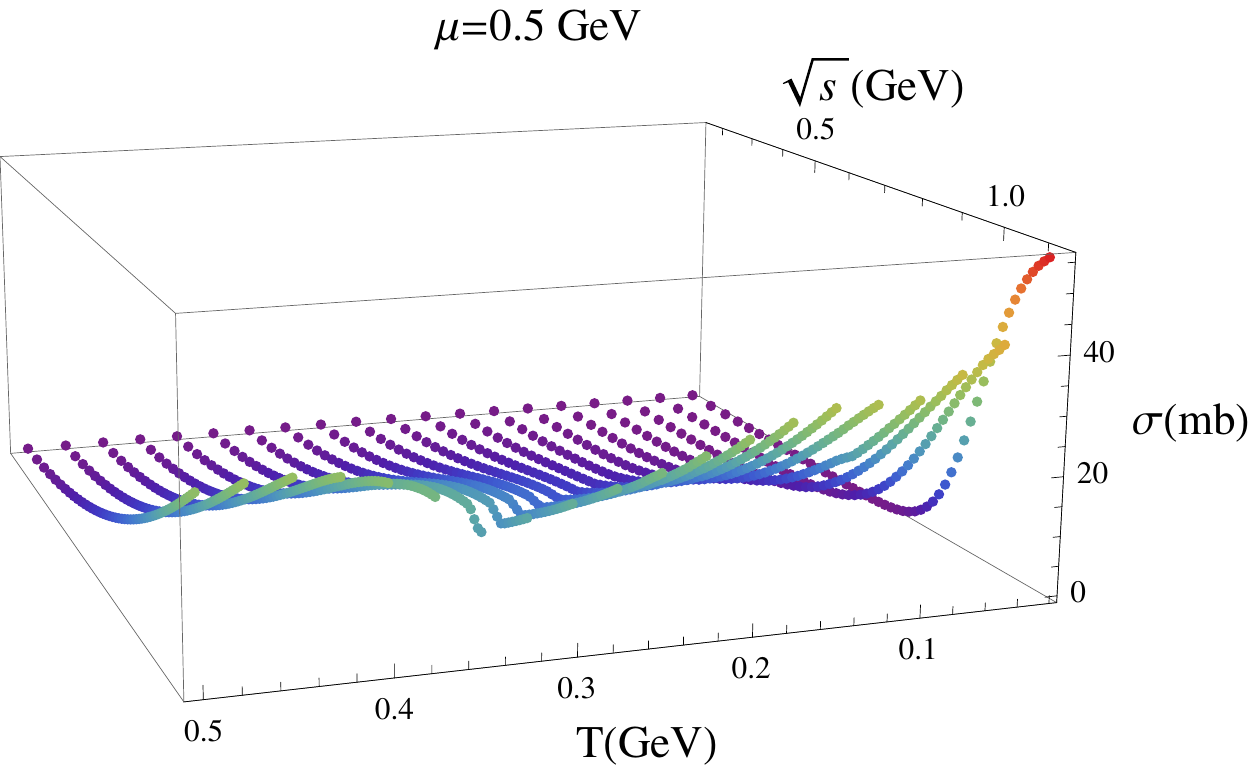}
\end{minipage}
\caption{\label{fig:uubar}$u\bar u \rightarrow u\bar u$ cross section as a function of the temperature and of the center of mass energy for different chemical potentials $\mu$.}
\end{figure}
 $u$, $\bar{u}$ and $v$, $\bar{v}$ are the Dirac spinors. $\delta_{c3,c4}$ impose the color conservation. The symbolic expression 
 $T \mathcal{D}_i^{S/P} T$ stands for the exchange of all possible scalar and pseudoscalar mesons, respectively. We limit us in this calculation to the exchange of color neutral mesons and the $s$-channel is taken
in first order in $N_c$. 
The square of the matrix element, averaged over spin and color in  the entrance and summed over in the exit channel,
\begin{equation}
\frac{1}{4 N_c^2}  \sum_{s,c} |\mathcal{M}_{total}|^2 
\end{equation}
 gives the cross section
\begin{equation}
\frac{\mathrm{d}\sigma}{\mathrm{d}t} = \frac{1}{16 \pi [s-(m_1+m_2)^2][s-(m_1-m_2)^2]}\frac{1}{4N_c^2} \sum_{s,c} |\mathcal{M}_{total}|^2.
\end{equation} 
In a thermal heat bath some of the exit states are partially blocked. Introducing the Pauli blocking factors we obtain
\begin{equation}
\sigma = \int \frac{\mathrm{d}\sigma}{\mathrm{d}t} (1-f^{\pm}(E_3,T,\mu_3))(1-f^{\pm}(E_4,T,\mu_4)) \mathrm{d}t .
\end{equation}
The integration limits for
 $t$ are $-(s- \sum_{i} m_i^2)$ and  0.
In order to calculate the transition amplitudes one has to know the mesons which can be exchanged and the corresponding flavor factors.
The exchanged mesons are either scalars $\sigma_\pi$, $\sigma_K$, $\sigma$ and $\sigma'$ or pseudoscalars  $\pi$ ,$K$, $\eta$ and $\eta'$. They are displayed in Table~\ref{tab:qqbar}.
The Gell-Mann matrices for the different flavors are  
\begin{equation}
\begin{cases}
\pi^0, \lambda_3 \\
\pi^\pm, \frac{1}{\sqrt{2}}(\lambda_1 \pm i \lambda_2) \\
K^0,\bar{K}^0, \frac{1}{\sqrt{2}}(\lambda_6 \pm i \lambda_7) \\
K^\pm,\frac{1}{\sqrt{2}}(\lambda_4 \pm i \lambda_5).
\end{cases} 
\end{equation}
To calculate which mesons can be exchanged one has to calculate the corresponding flavor matrices. As an example we calculate $ud \rightarrow ud$ with the exchange of a $\pi^{0}$ in the $t$-channel. Here one finds
\begin{equation} 
\begin{split}
\bar{u} \lambda_3 u \times \bar{d} \lambda_3 d= 
\begin{pmatrix} 1 & 0  & 0 \end{pmatrix} \begin{pmatrix} 1 & 0 & 0 \\ 0 & -1 & 0 \\ 0 & 0 & 0 \end{pmatrix} \begin{pmatrix} 1 \\ 0\\ 0 \end{pmatrix} \times
\begin{pmatrix} 0 & 1  & 0 \end{pmatrix} \begin{pmatrix} 1 & 0 & 0 \\ 0 & -1 & 0 \\ 0 & 0 & 0 \end{pmatrix} \begin{pmatrix} 0 \\ 1\\ 0 \end{pmatrix} = -1.
\end{split}
\end{equation}
\begin{figure}[htp]%
\centering
\begin{minipage}{0.49\textwidth}
\centering
\includegraphics[scale=0.5]{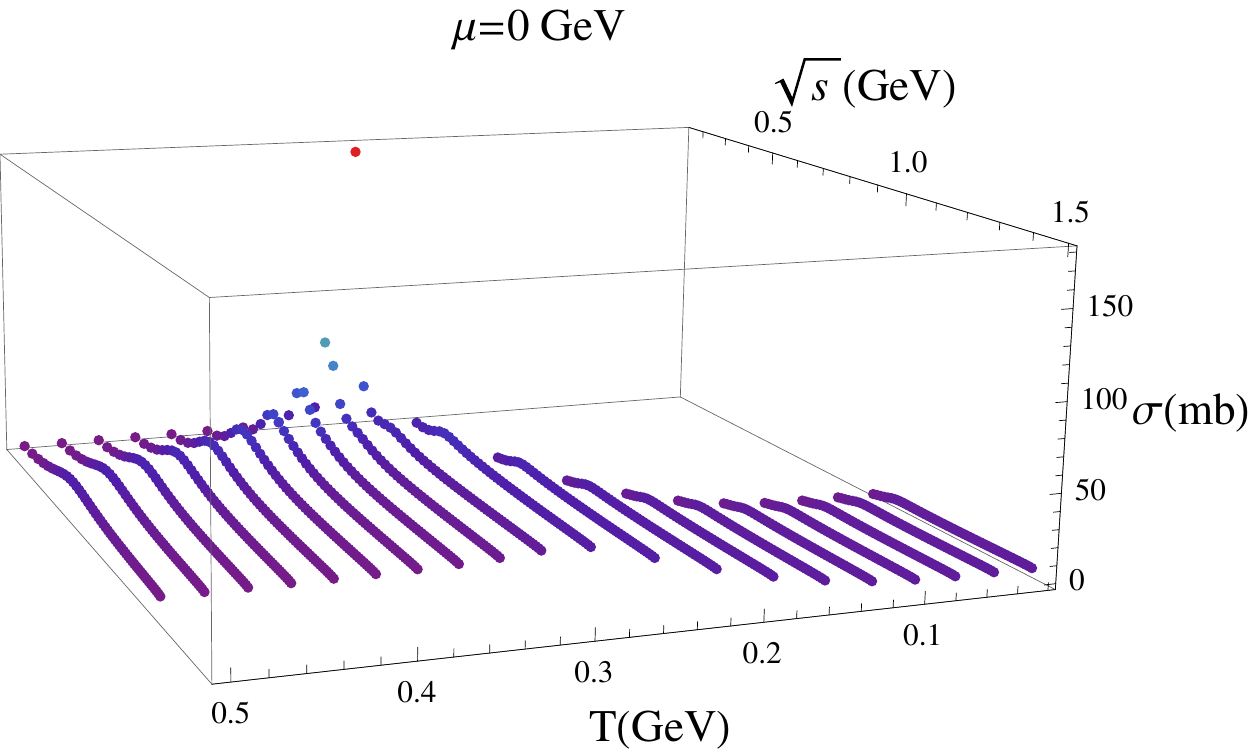}
\end{minipage}
\hfill
\begin{minipage}{0.49\textwidth}
\includegraphics[scale=0.5]{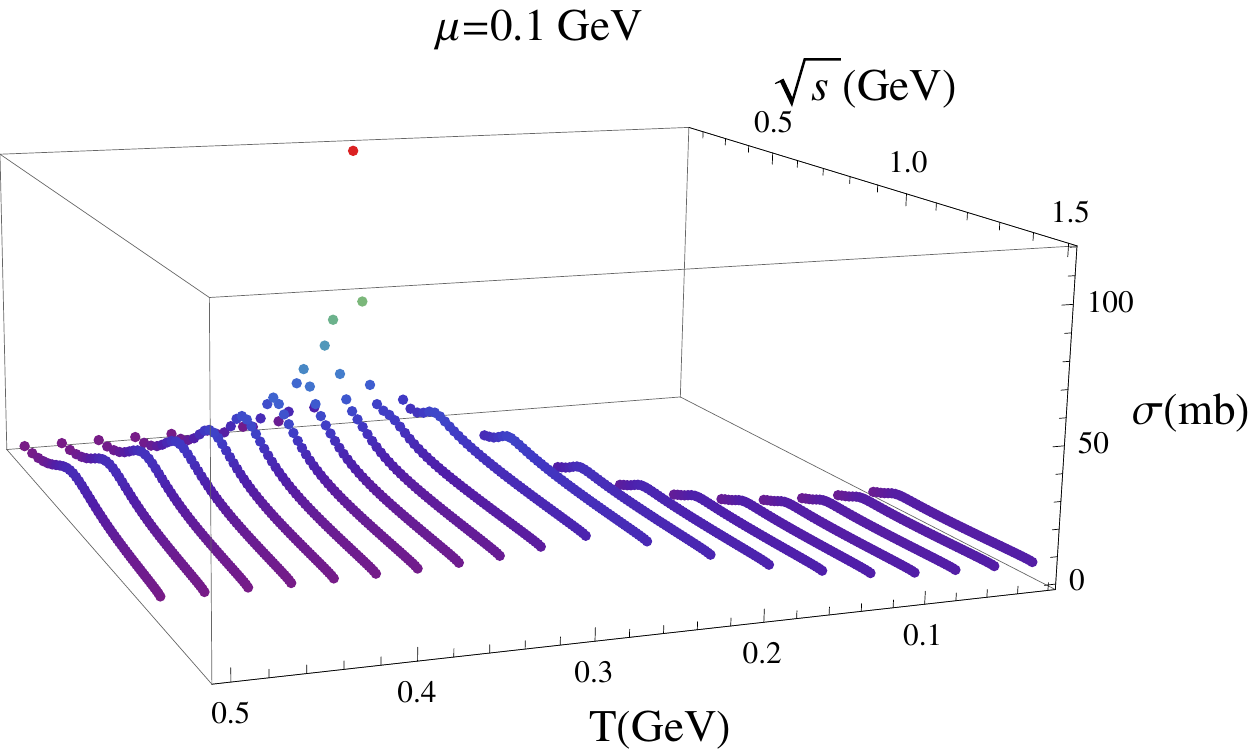}
\end{minipage}
\hfill
\begin{minipage}{0.49\textwidth}
\includegraphics[scale=0.5]{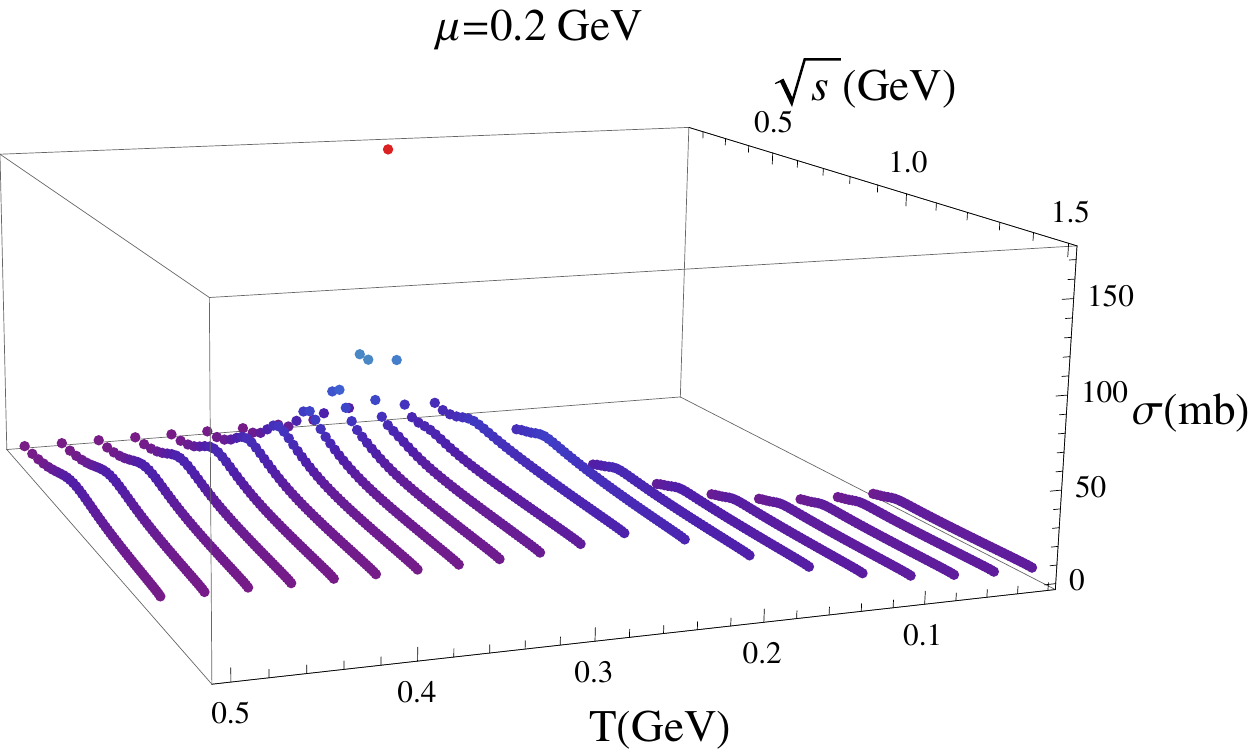}
\end{minipage}
\hfill
\begin{minipage}{0.49\textwidth}
\includegraphics[scale=0.5]{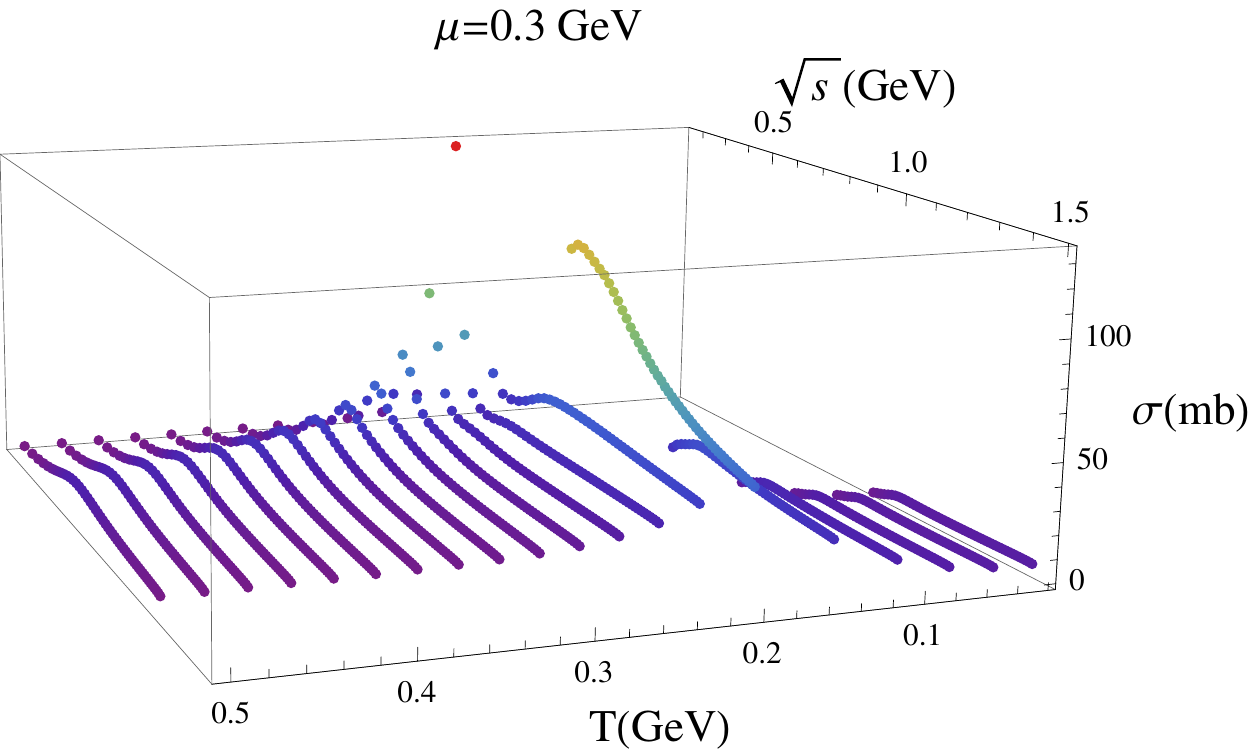}
\end{minipage}
\hfill
\begin{minipage}{0.49\textwidth}
\includegraphics[scale=0.5]{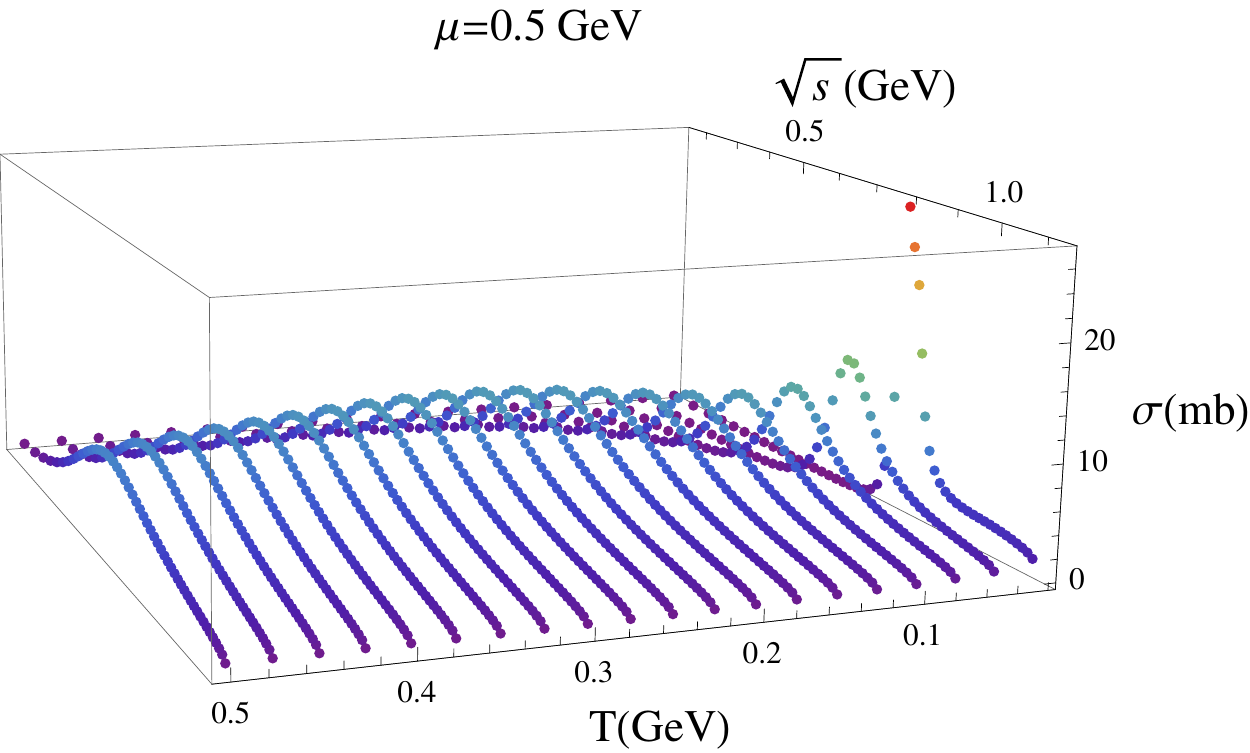}
\end{minipage}
\caption{\label{fig:udbar}Cross section for $u\bar{d} \rightarrow u\bar{d}$ as a function of the temperature T and the center of mass energy  $\sqrt{s}$ for different chemical potentials $\mu$.}
\end{figure}
Hence a $\pi_0$ can be exchanged, whereas for a $\pi^+$ we find zero. If we apply this for all possible combinations we find the mesons of Table~\ref{tab:qqbar}.
\begin{table}
\centering
\caption{\label{tab:qqbar}The mesons which can be exchanged in the different channels for  $q\bar{q}\rightarrow q\bar{q}$}
\begin{tabular}{|c | c | c |}
\hline
Process & Mesons exchanged in the s-channel & Mesons exchanged in the  t-channel \\ \hline
$u\bar{u} \rightarrow u\bar{u}$ &$ \pi, \eta, \eta',\sigma_{\pi}$,$ \sigma, \sigma' $&$ \pi, \eta, \eta',\sigma_{\pi}, \sigma, \sigma' $\\ \hline
$u\bar{u} \rightarrow u\bar{d}$ & $\pi, \eta, \eta',\sigma_{\pi}$, $\sigma, \sigma'$ & $ \pi, \sigma_{\pi} $\\ \hline
$u\bar{d} \rightarrow u\bar{d}$ & $   \pi, \sigma_{\pi} $ & $ \pi, \eta, \eta',\sigma_{\pi}$,$ \sigma, \sigma'  $ \\ \hline 
$u\bar{s} \rightarrow u\bar{s}$ &$ K, \sigma_K$ & $\eta, \eta',  \sigma, \sigma' $ \\ \hline
$u\bar{u} \rightarrow s\bar{s}$ & $\eta, \eta',  \sigma, \sigma'  $ & $K, \sigma_K $ \\ \hline
$s\bar{s} \rightarrow u\bar{u}$ & $ \eta, \eta',  \sigma, \sigma' $ & $K, \sigma_K $ \\ \hline
$s\bar{s} \rightarrow s\bar{s}$ &$\eta, \eta',  \sigma, \sigma' $ & $\eta, \eta',  \sigma, \sigma'$ \\ \hline
\end{tabular}
\end{table}

\section{Results}
The $q\bar q$ cross section has a quite different behavior as compared to the $qq$ cross section due to the $s$-channel contribution. The $qq$ cross section is always $\le $20 mb. In the $s$ channel it happens that
the incoming $q\bar q$ pair is in resonance with the meson which it produces. In this case the denominator of the meson propagator becomes  small  and hence the matrix element large. This resonance between the
incoming $q\bar q$ pair appears close to the temperature where the mass of the meson is close to the sum of the masses of the two valence quarks. Therefore the peak of the cross section moves to lower
temperatures when the chemical potential increases. 

\subsection{$u\bar{d} \rightarrow u\bar{d}$ and $u\bar{u} \rightarrow u\bar{u}$ }
Close to  the critical chemical potential, $\mu_{crit}$, (see Table~\ref{tab:para}) the cross section becomes maximal being around 100  mb for the $u \bar u$ channel at the grid points which we calculated. At larger
chemical potentials the cross sections are smaller again and arrive at a maximal value of 25 mb at $\mu = 0.5$ MeV (see Fig.\ref{fig:uubar}). For even higher chemical potentials the cross section is reduced to a couple of millibarn
 because then a transition between the plasma and the hadronic world does not exist anymore in the NJL approach. There we are for all temperatures in the deconfined phase.    

If one compares the $u\bar u$ and $u\bar d$ elastic cross sections, Figs. \ref{fig:uubar} and  \ref{fig:udbar}, one sees that the latter is about two to four times larger. This is a consequence of the flavor factor 
which doubles the $s$-channel contribution and gives a relative minus sign to the $t$-channel contribution. The form of the cross sections are rather similar, only at large values of $\sqrt{s}$ the fact that the $\eta$ and
$\eta'$ mesons are not allowed in the $s$ channel of $u\bar d$ makes the $u\bar u$ cross section relatively larger.  
\begin{figure}[htp]%
\centering
\begin{minipage}{0.49\textwidth}
\centering
\includegraphics[scale=0.5]{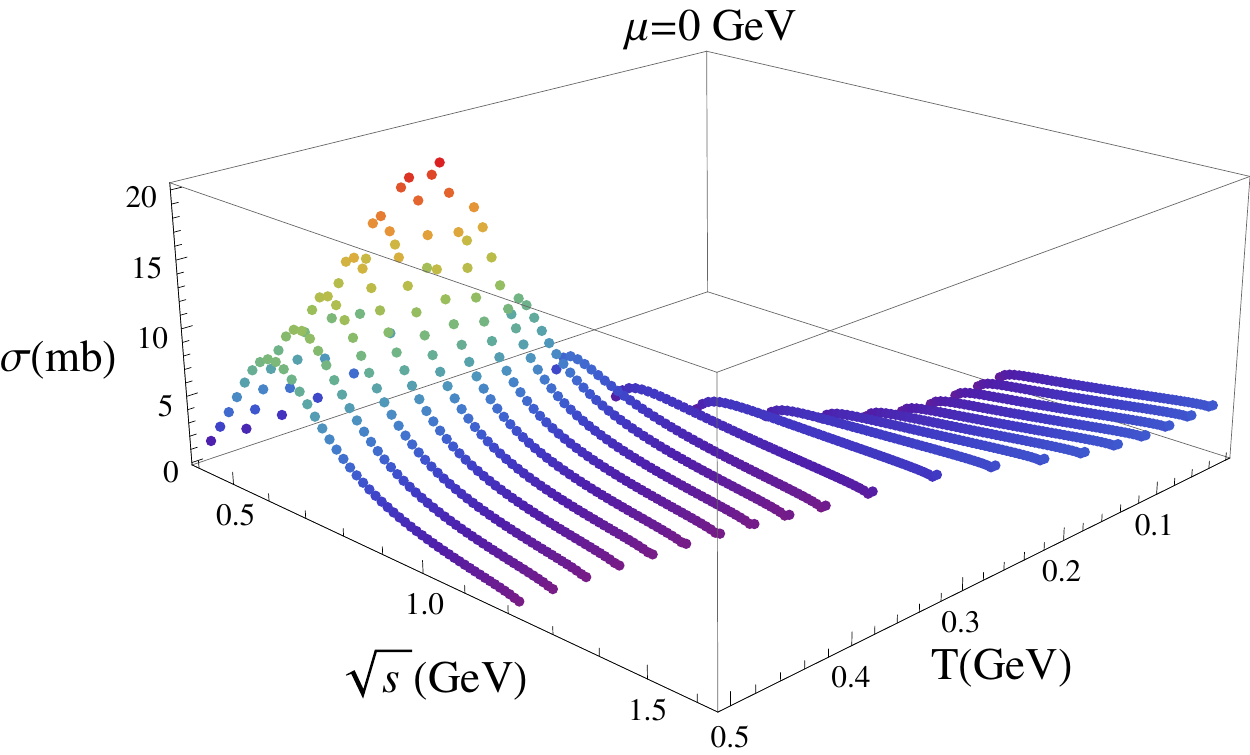}
\end{minipage}
\hfill
\begin{minipage}{0.49\textwidth}
\includegraphics[scale=0.5]{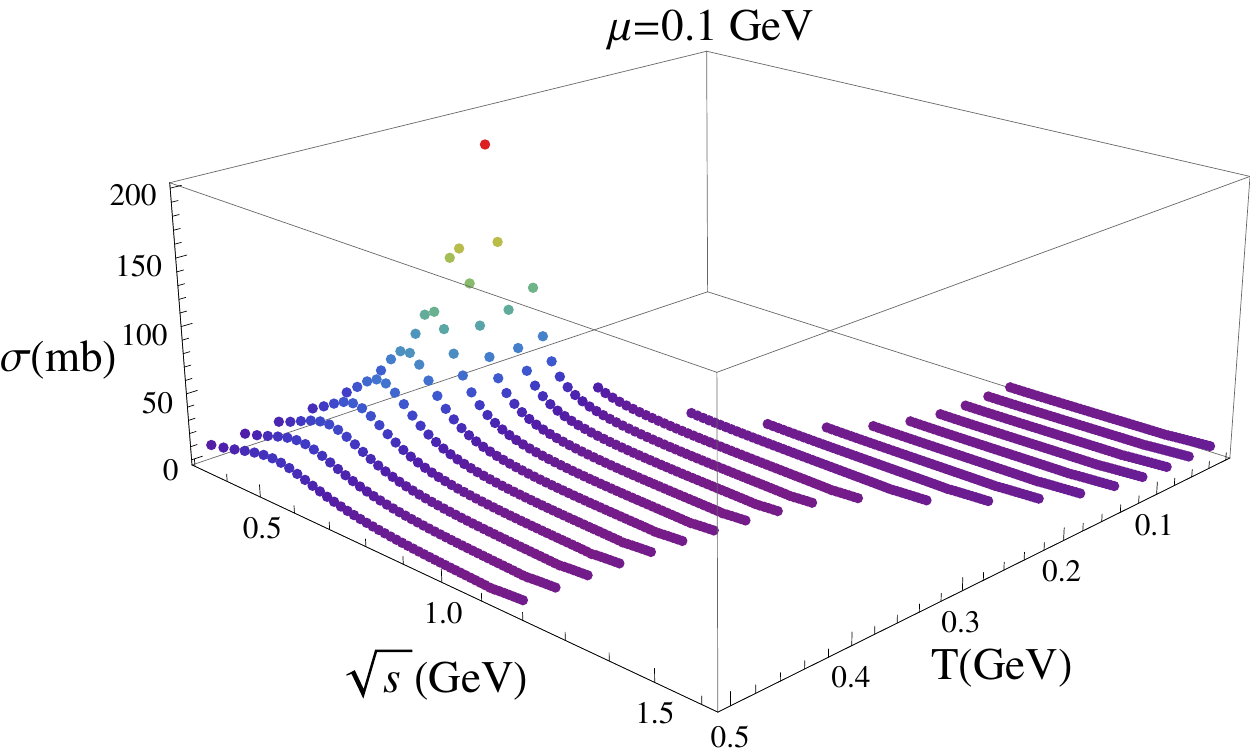}
\end{minipage}
\hfill
\begin{minipage}{0.49\textwidth}
\includegraphics[scale=0.5]{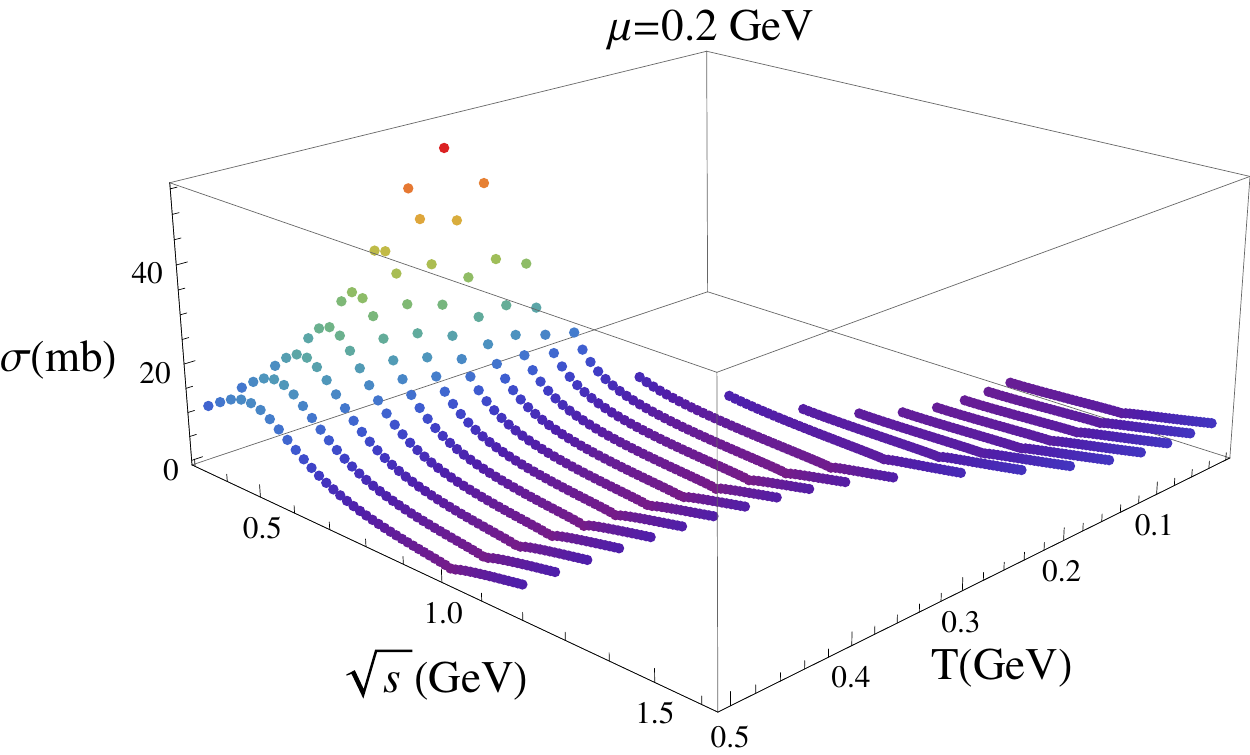}
\end{minipage}
\hfill
\begin{minipage}{0.49\textwidth}
\includegraphics[scale=0.5]{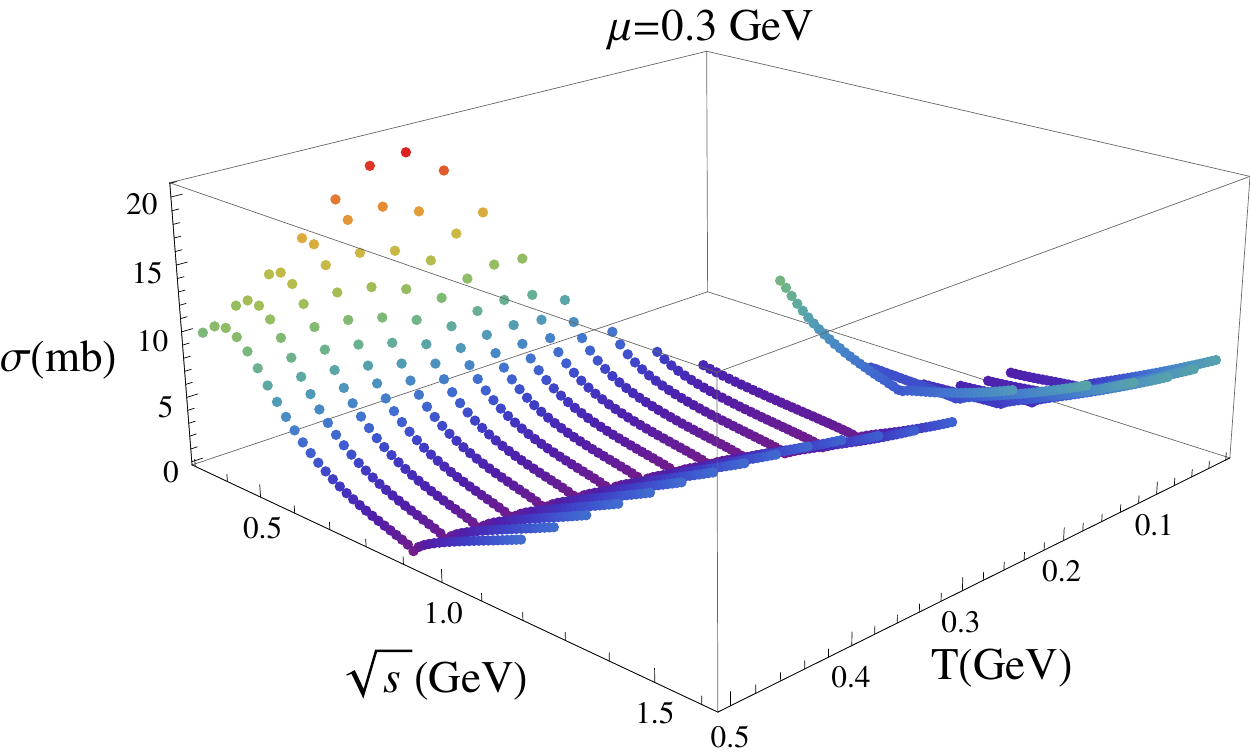}
\end{minipage}
\hfill
\begin{minipage}{0.49\textwidth}
\includegraphics[scale=0.5]{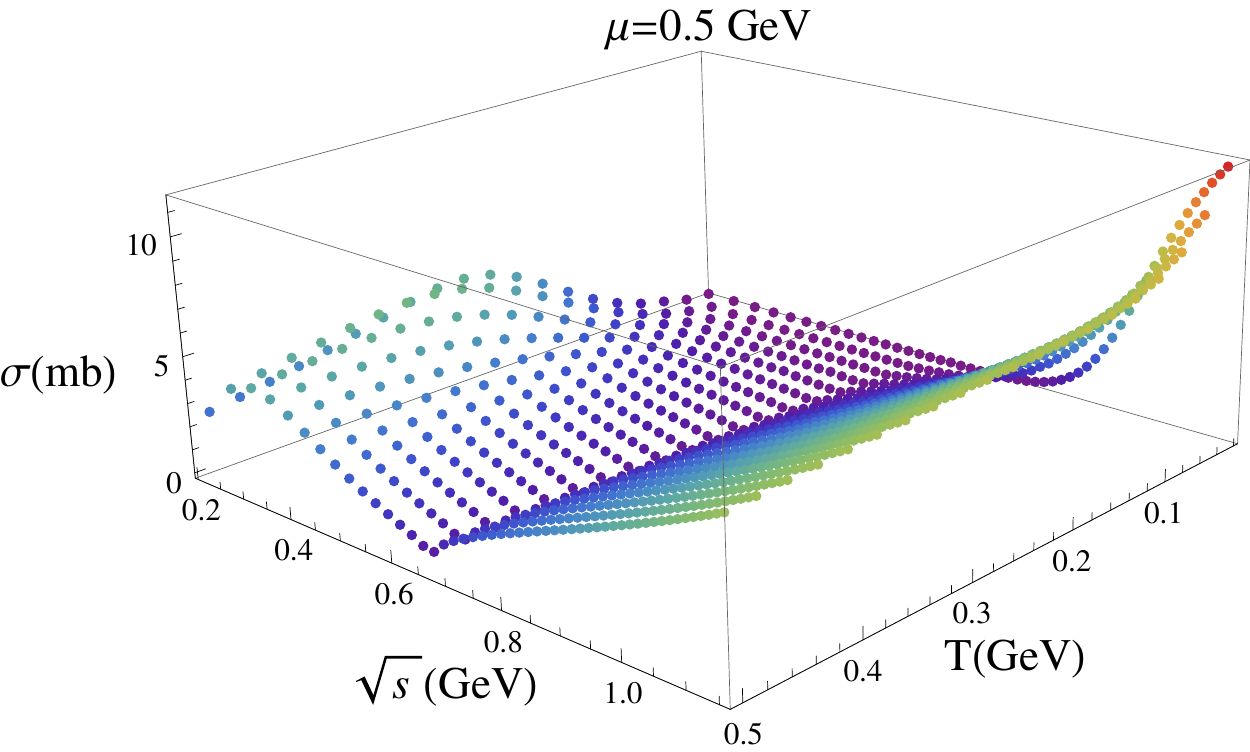}
\end{minipage}
\hfill
\caption{\label{fig:usbar}Cross section for $u\bar{s} \rightarrow u\bar{s}$ as a function of the temperature $T$ and the center of mass energy  $\sqrt{s}$ for different chemical potentials $\mu$.}
\end{figure}
\subsection{$u\bar{s} \rightarrow u\bar{s}$ }
The  $u\bar{s}$ cross section is displayed in Fig.~\ref{fig:usbar}. As compared to the other cross sections the maximum of the cross section is obtained at a higher temperature. This is a consequence that now in the
$s$ channel strange mesons are exchanged which have a mass well above the mass of the quarks in the entrance channel. Therefore more energetic particles are needed to be resonant with the exchanged meson.

\section{Conclusions}
This calculation of the cross sections at finite $\mu$ is the first step towards a transport theory based on the NJL Lagrangian, for finite chemical potentials, extending the work of Ref.~\cite{Marty1,Marty2}. We see that also at a finite chemical potential close to the transition between the  partonic and hadronic world the elastic cross sections become very large and therefore are very effective to equilibrate the system when it has expanded to the density where the phase transition takes place. Before the cross sections are too low to keep a local equilibrium. Consequently, for finite chemical potentials we expect the same generic behavior which we have observed at zero chemical potential, even if the numerical values of the cross sections differ in details. Therefore the NJL transport approach may be extended to a finite chemical potential to study how the change of the structure of the phase transition is seen in the observables.  
\section*{Acknowledgement}
This work has been funded by the program TOGETHER from R\'egion Pays de la Loire and EU Integrated Infrastructure Initiative HadronPhysics3 Project under Grant Agreement
n. 283286. JMTR thanks funding from a Helmholtz Young Investigator Group VH-NG-822 from the Helmholtz Association and GSI, and the Project FPA2013-43425-P from
Ministerio de Ciencia e Innovaci\'on (Spain).

\section*{References}
\medskip

\smallskip
\end{document}